\documentclass[debug,overfull]{epl}
\usepackage{amsmath}
\usepackage{graphics}

\title{Stretched exponential dynamics in a chain of coupled\\ chaotic
oscillators.}
\shorttitle{Stretched exponential dynamics}

\author{E. R. Hunt\inst{1}\thanks{E-mail: hunt@helios.phy.ohiou.edu}, P. M. Gade\inst{1} and Normand
Mousseau\inst{1,2}\thanks{E-mail:normand.mousseau@umontreal.ca}}

\institute{
 \inst{1} Department of Physics and Astronomy and CMSS, Ohio University,
Athens, OH 45701, USA \\
\inst{2} D\'epartement de phy\-si\-que,
Uni\-ver\-sit\'e de Mont\-r\'eal, C.P. 6128, Succ. Centre-ville, Montr\'eal
(Qu\'ebec) Canada H3C 3J7.
}


\begin{document} 

\maketitle

\begin{abstract} 
We measure stretched exponential behavior,
$ \exp \left( -(t/\tau_0)^\beta \right)$,
over many decades in a one-dimensional array of coupled
chaotic electronic elements just above a crisis-induced intermittency
transition. There is strong spatial heterogeneity and individual sites display a
dynamics ranging from near power law ($\beta=0$) to near exponential ($\beta=1$)
while the global dynamics, given by a spatial average, remains  stretched
exponential. These results can be reproduced quantitatively with a
one-dimensional coupled-map lattice and thus appear to be system independent. In
this model, local stretched exponential dynamics is achieved without frozen
disorder and is a fundamental property of the coupled system. The
heterogeneity of the experimental system can be reproduced by introducing
quenched disorder in the model. This suggests that the stretched exponential
dynamics can arise as a purely chaotic phenomenon.
\end{abstract}

Stretched exponential relaxation is almost ubiquitous in complex systems. It is
found in glasses~\cite{angell96}, spin glasses~\cite{spinglasses}, polymers,
high-dimensional cellular automata~\cite{mousseau96}, random
networks~\cite{krapivsky00} coupled chaotic oscillators~\cite{tsironis96,ngai99}
and others~\cite{frogs}. For many of these problems, however, the microscopic
origin of this dynamics is difficult to establish both experimentally and
theoretically.  Understanding the underlying causes for stretched exponential
relaxation remains, therefore, a central goal in the study of the dynamics of
complex systems~\cite{tracht98,israeloff00,cugliandolo00}, as witnessed by the
flurry of theoretical, experimental and numerical results that have
appeared on the subject in the last few years
~\cite{glotzer98,rabani99,weeks00,cugliandolo00}.

In the last 20 years, many physical processes have been understood
in terms of nonlinear dynamics and chaos, these concepts often going beyond the
reach of standard equilibrium statistical physics.  Here, we present
experimental results showing a stretched exponential dynamics, $\exp \left[
-(t/\tau_0)^\beta \right]$, in a simple one-dimensional chaotic system comprised
of a chain of chaotic electronic oscillators.  In particular, we find that in
the vicinity of the transition to spatiotemporal chaos, individual site dynamics
shows this behavior over six orders of magnitude and that the $\beta$'s
associated with individual site range from 0.8 (nearly exponential) to 0.2
(nearly power law) depending on the activity of the site. Furthermore, the
global dynamics of the networks also shows a stretched-exponential dynamics with
a $\beta$ having an intermediate value.  These results can be reproduced using a
1D coupled-map model which also displays a transition to spatio-temporal chaos.
Close to that transition, the model shows uniform stretched exponential
dynamics. We recover the heterogeneous dynamics of the experiment by introducing
quenched disorder in the model. These results suggest that the stretched
exponential dynamics can arise as a purely chaotic phenomenon.

\begin{figure}
\onefigure[width=12cm]{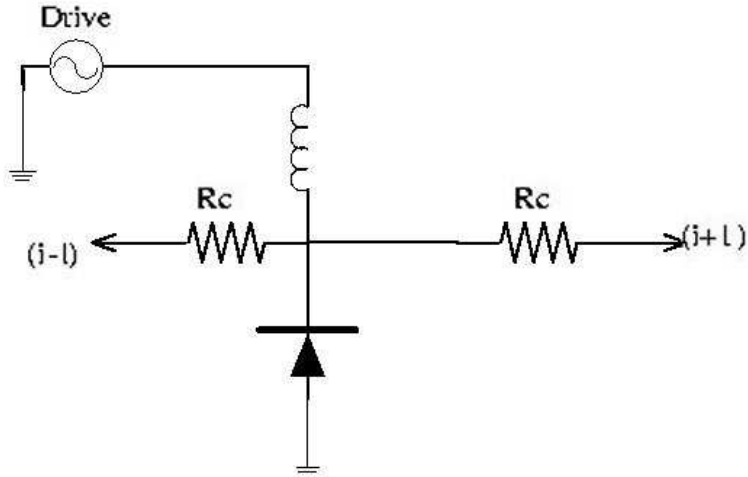}
\caption{Schematic diagram of a single diode-resonator and its coupling to its
nearest neighbors in a one-dimensional arrangement.}
 \label{fig:setup}
\end{figure}

We study the behavior of a one-dimensional chain of 256 coupled diode-resonators
with periodic-boundary conditions.  Figure \ref{fig:setup} shows the schematic
diagram of the basic blocks forming the one-dimensional array. A diode resonator
is comprised of the series combination of an inductor (30 mHy) and a rectifier
type pn junction diode (1N1004) and is driven by a sinusoidal source (100 kHz);
it follows the period-doubling route to chaos as the ac drive is
increased~\cite{hunt}.  There are fairly large differences in the individual
resonators; for example, the drive voltage for a given feature, such as the
period-3 fixed point, varies by as much as 20 \%. Diffusive coupling between the
elements is provided by resistors as shown and periodic boundary conditions are
used.  This one-dimensional setup has been previously studied with
unidirectional coupling~\cite{johnson96} and under the conditions of stochastic
resonance~\cite{locher98b}.

As the drive voltage is increased, the coupled system shows a kink-forming route
to spatiotemporal chaos.  At low drive the elements period double nearly
synchronously with no or very little current in the coupling resistors.
Increasing the drive, the system reaches a two-band chaotic state in which all
elements are in the same band. Following this, coarse grained period-2 phase
kinks (two-band structures in which neighboring elements are in opposite bands)
begin to appear~\cite{johnson96}. A kink tends to make the system less chaotic
since there is now appreciable power dissipated in the coupling resistors and
the available phase space is reduced. With increasing drive voltage, the number
of kinks increases until a regular or a near regular kink lattice is formed with
a lattice constant function of the coupling resistors.  At this point each
element displays two-band chaos with the phase alternating around the chain.
Further increasing the drive, the two chaotic bands merge into one, in an
example of an attractor-merging crisis in a high-dimensional spatiotemporal
system~\cite{ott93}.  The critical drive voltage V$_c$ is where the two
attractors just touch.  Well above V$_c$ the system displays full spatiotemporal
chaos. Just above V$_c$ regions of the system still spend long periods of time
trapped in the old two-band attractor before switching to the new single
attractor for a time and then back again.  In a single chaotic oscillator the
intermittent switching between two attractors is an example of crisis-induced
intermittency and the trap time displays exponentially distributed switching
times~\cite{ott93}. As discussed below, crisis-induced intermittency also exists
in coupled system but the trap time at that point is now dominated by stretched
exponentials.

\begin{figure} 
\hspace*{-1cm} 
\onefigure[width=12cm]{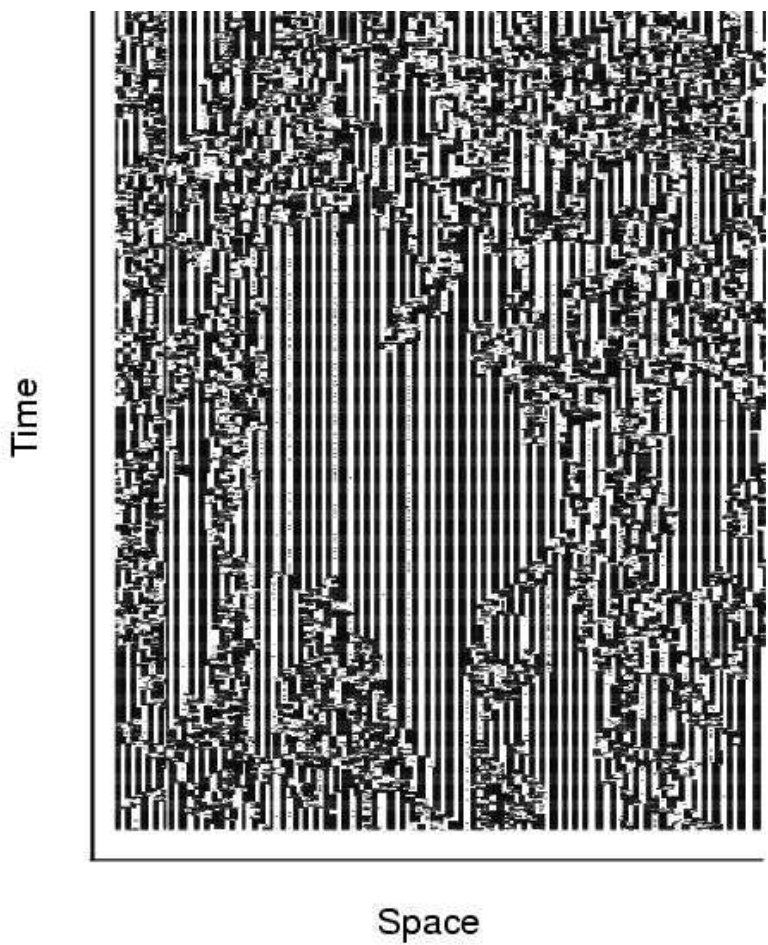}
\vspace*{-2cm}
 \caption{ Time series of the 256 diodes in the intermittent
state, just above $V_c$. Since the system displays a fundamental period-two
oscillation in time below $V_c$,  we show  the state of the diodes  at even
time-steps in a binary representation.}
\label{fig:traps_exp} 
\end{figure}

Figure \ref{fig:traps_exp} shows a 4096-cycle time series for the 256 elements.
The two phases of the two-band attractor are represented as black and white, and
only every other time step is displayed. Note the creation, annihilation and
diffusion of the boundaries of the two-band attractor.  With the coupling resistors used, 
150 k$\Omega$, the coarse-grained spatial periodicity of the lattice is about 4 sites.

\begin{figure}
\onefigure[width=12cm]{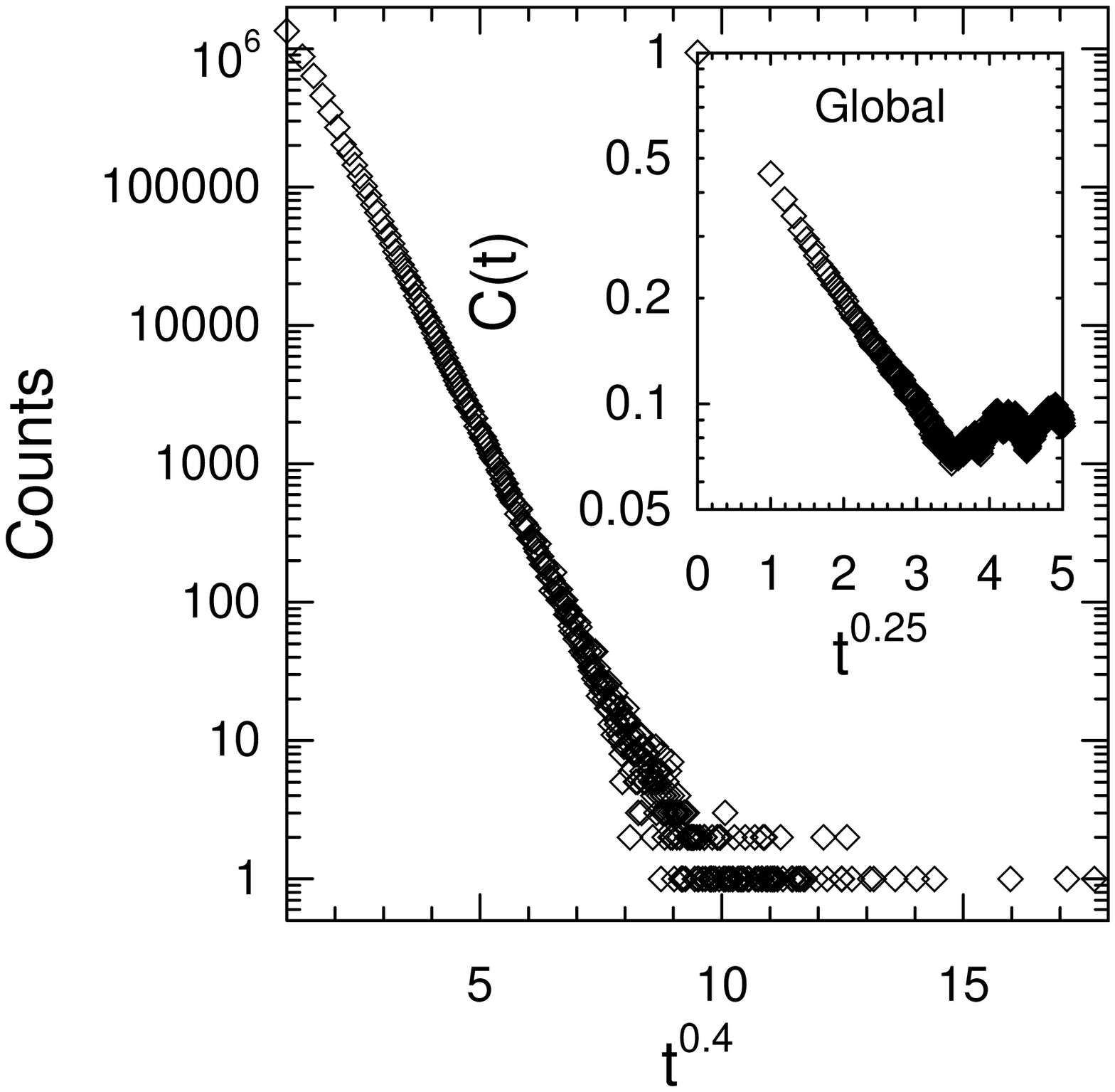}
\caption{ Distribution of trap times measured over a single site  over a 10-hour
period by interval of 1 ms.  The raw data are shown to fit to a stretched
exponential with $\beta=0.4$. The apparent tail at longer time is due to the
discrete nature of the measurement and could be redressed by more statistics.
Inset: Time auto-correlation function averaged over all 256 sites for a run of
of 2048 timesteps of 64 drive cycles each. The data fits a stretched exponential
with $\beta=0.25$.} 
\label{fig:distr_exp} 
\end{figure}

In Fig.~\ref{fig:distr_exp}, we plot the distribution of trap times, i.e., how
long a site remains in the two-band phase. This distribution is measured on a
single site over $10^9$ cycles, or 10 hours of experimental measurements.  This
distribution is very well approximated by a stretched exponential with an
exponent $\beta = 0.40 \pm 0.05$ and $\tau_0 = 27 \pm 10 $ drive cycles.  In
glasses and other systems, this behavior is associated with relaxation
processes. These processes are described by a time autocorrelation function
averaged over the entire system. As shown in the inset of
Fig.~\ref{fig:distr_exp}, our system also shows a global stretched-exponential
dynamics; the flattening of the auto-correlation function at longer time is due
to the limited statistics~\cite{setup}.

\begin{figure} 
\onefigure[width=12cm]{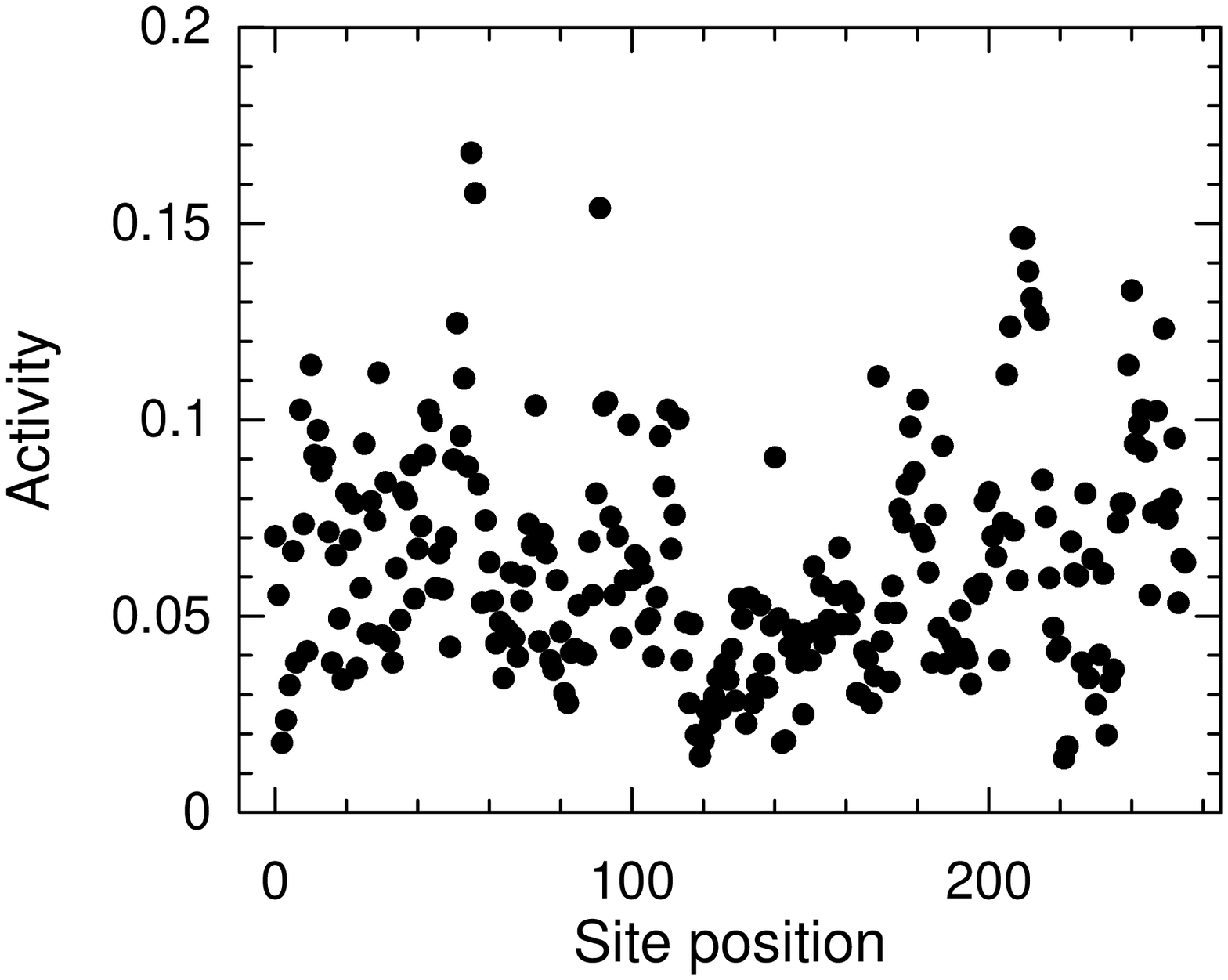}
\caption{Activity (fraction of the time steps where a site changes phase) as a
function of site number for the 256 coupled diode-resonators.}
\label{fig:activity}
 \end{figure}

We can address directly the question of heterogeneities in our setup. Figure
\ref{fig:activity} shows the activity associated with each site.  Activity is
defined as the number of times a site changes state normalized by the number of
time steps. The activity is a measure of the local dynamics of the sites: very
active sites typically show a dynamics close to that of the chaotic regime while
the slow sites display a trap-time distribution more in line with the frozen
regime. Mostly determined by the value of the local chaotic threshold, the
variations in activity are essentially uncorrelated spatially as can be seen in
Fig. ~\ref{fig:activity}.  Spatial correlations, measured as a two-point
function on the lattice, show an exponential decay, $\exp(- r/r_0)$, with $r_0
\simeq 6$ sites, in the intermittent phase, close to the basic spatial
periodicity just below crisis.

\begin{figure} 
\onefigure[width=12cm]{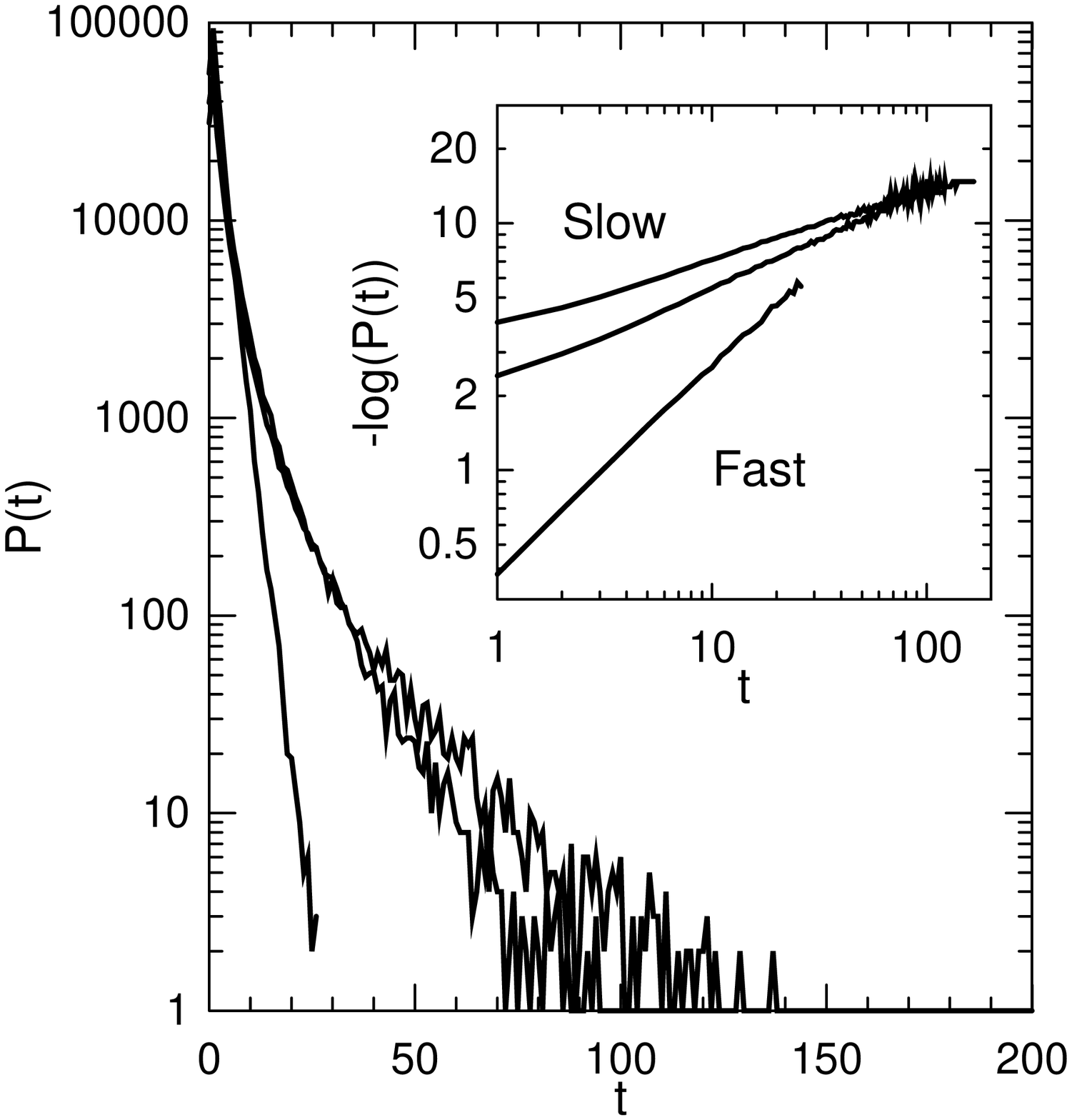}
\caption{Trap-time distribution for 3 diode-resonators with different activity,
as defined in Fig. \protect\ref{fig:activity}. The main figure shows a
log-normal plot of the distribution for all sites. In inset, we show a log-log
vs. log plot for the three sites; the curves were translated vertically for
better display. From left to write, in the main panel (and bottom to top in the
inset), the corresponding $\beta$ is, respectively, 0.8, 0.4 and 0.3. }
\label{fig:individuals}
 \end{figure}

The level of activity for a given site is also reflected in the overall dynamics
of this site.  Figure \ref{fig:individuals} shows the trap-time distribution for
three sites with different level of activity. All three sites are consistent
with stretched exponential with exponents $\beta=0.8$ for the ``hotter'' site,
$\beta=0.4$ and $\beta=0.3$, for the ``cooler'' sites.  Faster sites can be
equated with an effective higher driving voltage and a dynamics moving towards a
pure exponential ($\beta=1$); slower sites, on the other hand, display stretched
exponential with lower exponent, and can even have a power-law correlation
($\beta=0$), depending on the external driving. For a big enough system, we
expect to find the whole range of $\beta$ at any driving in this intermittent
phase and see the distribution of exponent change with driving, with the value
of the global $\beta$ somewhere in between the two extremes contrary to what was
proposed recently by Richert and Richert~\cite{richert98}.

As is often the case for glasses, where the exponent $\beta$ varies as a
function of temperature, $\beta$ here changes with the driving voltage. 
 At low
external driving, $\beta$ is zero or nearly zero and the system is either
totally frozen or showing a power-law distribution of trap times. As the driving
is increased, $\beta$ also increases relatively smoothly to $\beta=1.0$, for the
single site chosen, and to about 0.5, on average, before the full system become
completely chaotic.  The relaxation time scale, defined by $\tau_0$, follows
$\beta$ rather closely, roughly doubling in the driving range of 3 to 3.2. The
overall time scale associated with the traps is therefore of the order of 100
drive cycles, with trap times reaching many times this value during the
measurement.

\begin{figure}
\onefigure[width=12cm]{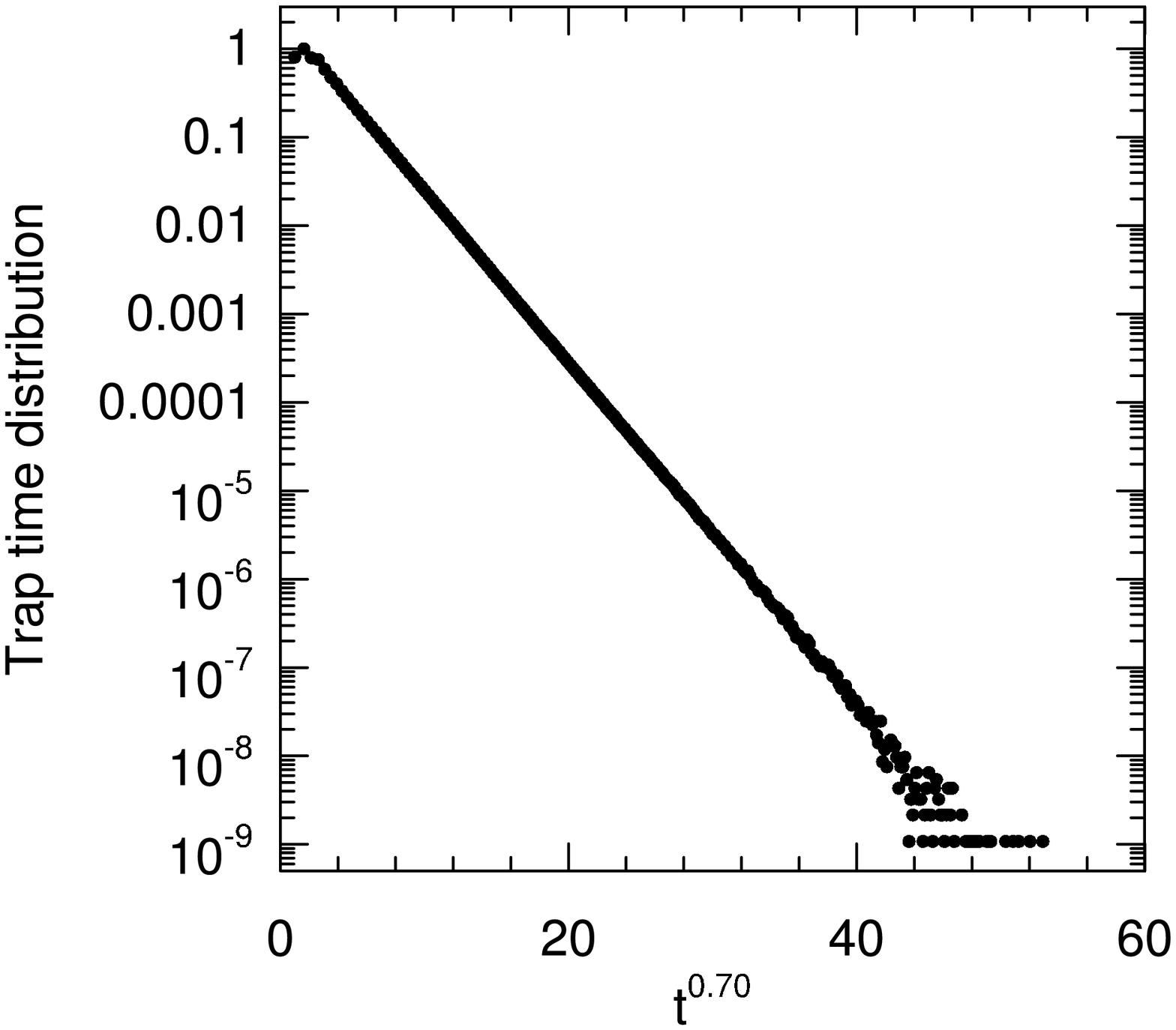}
\vskip 10pt
 \caption{ Spatially-averaged trap-time distribution for a
256-coupled-map lattice with $r=1.890$ and $\alpha=0.25$, run for 100 million
iterations. Each site displays the same stretched exponential relaxation, with
$\beta=0.70 \pm 0.05$, and we show the spatial average for better
statistics.} 
\label{cmnodisorder} 
\end{figure}

Surprising results are obtained in the simulation of a simple one-dimensional
coupled-map lattice.  This model follows Kaneko~\cite{kaneko} and Johnson {\it
et al.}~\cite{johnson96} and is composed of a chaotic map, $f(x)= 1-rx^2$,
coupled to its nearest neighbors in a one-dimensional chain 
\begin{equation}
f_i(t+\Delta t) = (1-\alpha) f_i(t) + \frac{\alpha}{2} \left[ f_{i-1}(t) +
f_{i+1}(t) \right]; 
\end{equation} 
$\alpha$ is the strength of the coupling
between the logistic maps. The model, like the experimental set-up, shows an
attractor-merging crisis with crisis-induced intermittency  in the vicinity of
the transition to spatiotemporal chaos. Below the transition each site is in a
coarse-grained period-two state. The system also shows spatial periodicity with
alternating groups of sites in a coarse-grade period-2 as described above
for the experimental set-up.  The
size of these groups depends on $\alpha$. We choose $\alpha=0.25$ for a
spatial wavelength of 4 sites, similar to the experiment.  Just above the crisis the
system gets trapped in the two-band attractor in local regions for a while
before going chaotic again. While for an uncoupled system the times between
switches have a long-time exponential distribution~\cite{ott93}, Fig.
~\ref{cmnodisorder} shows that the trap-time distribution for this model follows
a stretched exponential with $\beta= 0.70\pm 0.05$ over the full distribution,
with all sites displaying the same distribution. The stretched exponential is
therefore not caused by disorder but is a fundamental property of the coupled system. As can be seen in
Fig.~\ref{fig:traps_exp} traps show a spatial extension that appears 
to stabilize them, biasing the distribution towards longer time.
observed in single oscillators. 

Disorder is necessary to induce the strong site to site variation seen in the
experimental setup; it can be introduced in the model by selecting a
site-varying $r_i$ taken from a flat distribution of $\delta$,  $r_i \in
[r-\delta, r+\delta ]$.   As in the experiment, the global dynamics is also
well-described by a stretched exponential even though the various sites show a
broad range in their relaxation dynamics. A fuller discussion of the simulation
results will be presented elsewhere~\cite{gade01}.

In conclusion, we have found stretched exponential dynamics in a
one-dimensional chain of chaotic diode resonators near the transition to
spatiotemporal chaos. While individual sites display a wide range of $\beta$'s,
the global time autocorrelation function also shows this
behavior~\cite{glasses}. As can be seen in Fig.~\ref{fig:traps_exp} traps can
be continuously created and distroyed. They are stabilized by their spatial
extension, which biases the trap-time distribution towards longer times.  Based
on a simple one-dimensional coupled map model, we show that this behavior is
not related to disorder. These results indicate that the stretched exponential
dynamics can be purely chaotic in origin. 

\acknowledgements 
We  acknowledge partial support from the NSF under grant
number DMR-9805848 and the Natural and Engineering Research Council of Canada
(NM) and the ONR (ERH). We thank Don Carter for the device used to measure trap
times. NM is a Cottrell Scholar of Research Corporation.

\bibliographystyle{prsty} 

\begin{thebibliography}{99}


\bibitem{angell96} M. D. Ediger, C. A. Angell, and S. R. Nagel, J. Phys. Chem.
{\bf 100}, 13200 (1996); C. A. Angell, J. Phys. Cond Mat. {\bf 12}, 6463 (2000).

\bibitem{spinglasses} K. H. Fischer and J. A. Hertz, ``Spin glasses'',
Cambridge, 1991.

\bibitem{mousseau96} N. Mousseau, J. Phys. A: Math. Gen. {\bf 29}, 3021 (1996).

 \bibitem{krapivsky00} P. L. Krapisvky,
S. Redner and F. Leyraz, Phys. Rev. Lett. {\bf 85}, 4629 (2000).

\bibitem{tsironis96} G. P. Tsironis and S. Aubry, Phys. Rev. Lett. {\bf 77},
5225 (1996).

\bibitem{ngai99} K. Yeung Tsang and K. L. Ngai, Phys. Rev. E {\bf 54}, R3067
(1996); K. L. Ngai and K. Yeung Tsang, ibid {\bf 60}, 4511 (1999).

\bibitem{frogs} M. Donsker and S.R.S. Varadhan, Commun. Pure Appl. Math. {\bf
28}, 525 (1975); ibid {\bf 32}, 721 (1979).

\bibitem{tracht98} U. Tracht, M. Wihlhelm, A. Heuer, H. Feng, K. Schmidt-Rohr,
and H. W. Spiess, Phys. Rev. Lett. {\bf 81}, 2727 (1998).

\bibitem{israeloff00} E. Vidal Russel and N. E. Israeloff, Nature {\bf 408}, 695
(2000) 

\bibitem{cugliandolo00} L. F. Cugliandolo and J. L. Iguain, Phys. Rev. Lett.
{\bf 85}, 3448 (2000).

\bibitem{glotzer98} S. C. Glotzer, N. Jan, T. Lookman, A. B. MacIsaac, and P. H.
Poole, Phys.  Rev. E {\bf 57}, 7350 (1998). 

\bibitem{rabani99} E. Rabani, J. D. Gezelter and B. J. Berne, Phys. Rev. Lett.
{\bf 82}, 3649 (1999).

\bibitem{weeks00} E. R. Weeks, J. C. Crocker, A. C. Levitt, A. Schofield, and D.
A. Weitz, Science {\bf 287}, 627 (2000).

\bibitem{schlesinger84} M. F. Schlesinger and E. W. Montroll, Proc. Natl. Acad.
Sci. U.S.A. {\bf 81}, 1280 (1984); R. G. Palmer, D. L. Stein, E. Abrahams, and
P. W. Anderson, Phys. Rev. Lett.  {\bf 53}, 958 (1984). 

\bibitem{richert98} R. Richert and M. Richert, Phys. Rev. E {\bf 58}, 779
(1998).

\bibitem{hunt} P. S. Linsay, Phys. Rev. Lett. {\bf 47}, 1349 (1981); J. Testa,
J. P\'erez and C. Jeffires, ibid {\bf 48}, 714 (1982); R. W. Rollins and E. R.
Hunt, ibid {\bf 49}, 1295 (1982).

\bibitem{johnson96} G. A. Johnson, M. L\"ocher, and E. R. Hunt, Physica D {\bf
96}, 367 (1996).

\bibitem{locher98b} M. L\"ocher, D. Cigna, E. R. Hunt, G. A. Johnson, F.
Marchesoni, L.  Gammaitoni, M. E. Inchiosa, A. R. Bulsara, Chaos {\bf 8},  604
(1998);  
M. L\"ocher, D. Cigna, E. R. Hunt, Phys. Rev.  Lett. {\bf 80}, 5212 (1998); M. L\"ocher, N.
Chatterjee, F. Marchesoni, W. L. Ditto, and E. R.  Hunt, Phys. Rev. E {\bf 61},
4954 (2000). 

\bibitem{ott93} R. W. Rollins and E. R. Hunt, Phys. Rev. A {\bf  29}, 3327
(1984); E. Ott, ``Chaos in Dynamical Systems'', Cambridge, 1993.

\bibitem{setup} The experimental setup does not allow us to accumulate as much
statistics when averaging over all sites compared to single-site time averages.

\bibitem{kaneko} K. Kaneko, {\it Theory and Applications of Coupled Maps
Lattices}, Wiley and Sons, New York (1993).

\bibitem{gade01}  P. M. Gade, N. Chatterjee, E. R. Hunt, and N. Mousseau, in
preparation.

\bibitem{glasses} The overall behavior of our experiment and model demonstrate
also a more complex  picture than what is normally presented in the theory of
glasses: the global stretched-exponential dynamics is generated not from a sum
of simple exponentials with varying time scale but from stretched exponential
distributions with $\beta$ ranging from zero to one. This raises the possibility that 
the local dynamics in glasses could be more complex that what is described in the
traditional heterogeneous model.


\end{thebibliography}

\end{document}